\documentclass[useAMS,usenatbib]{mn2e} 
\usepackage{graphicx,epsfig,psfig} 
\def\spose#1{\hbox to 0pt{#1\hss}} 
\def\simlt{\mathrel{\spose{\lower 3pt\hbox{$\mathchar"218$}} 
\raise 2.0pt\hbox{$\mathchar"13C$}}} 
\def\simgt{\mathrel{\spose{\lower 3pt\hbox{$\mathchar"218$}} 
\raise 2.0pt\hbox{$\mathchar"13E$}}}

\title[The ISM Structure of Simulated Disc Galaxies]{The Role of Feedback 
in Shaping the Structure of the Interstellar Medium} 
\author[Walker et~al.]{A.P. Walker$^{1}$,
B.K. Gibson$^{1,2}$,
K. Pilkington$^{1,2}$,
C.B. Brook$^{3,1}$, 
P. Dutta$^{4}$,\newauthor
S. Stanimirovi\'c$^{5}$,
G.S. Stinson$^{6}$, and
J. Bailin$^{7}$\\
$^1$Jeremiah Horrocks Institute,
University of Central Lancashire, Preston, PR1~2HE, UK\\
$^2$Institute for Computational Astrophysics, Dept of Astronomy \& Physics,
Saint Mary's University, Halifax, NS, B3H~3C3, Canada\\
$^3$Departamento de F\'{i}sica Te\'{o}rica, Universidad Aut\'{o}noma de
Madrid, E-28049 Cantoblanco, Madrid, Spain\\
$^4$National Centre for Radio Astrophysics, Post Bag 3, Ganeshkhind, Pune,
411~007, India\\
$^5$Department of Astronomy, University of Wisconsin, 475 North Charter St.,
Madison, WI, 53706, USA\\
$^6$Max-Planck-Institut f\"ur Astronomie, K\"onigstuhl 17, 69117
Heidelberg, Germany\\
$^7$Department of Physics \& Astronomy, University of Alabama,
Tuscaloosa, AL, 35487-0324, USA}

\begin{document} 
\date{Accepted: 3 March 2014} 
\pagerange{\pageref{firstpage}--\pageref{lastpage}} \pubyear{2014} 
\maketitle 
\label{firstpage} 

\begin{abstract} 
We present an analysis of the role of feedback in shaping the neutral 
hydrogen (HI) content of simulated disc galaxies. For our analysis, we have 
used two realisations of two separate Milky Way-like ($\sim$L$\star$) discs 
- one employing a conservative feedback scheme (MUGS), the other 
significantly more energetic (MaGICC).  To quantify the impact of these 
schemes, we generate zeroth moment (surface density) maps of the inferred HI 
distribution; construct power spectra associated with the underlying 
structure of the simulated cold ISM, in addition to their radial surface 
density and velocity dispersion profiles.  Our results are compared with a 
parallel, self-consistent, analysis of empirical data from THINGS (The HI 
Nearby Galaxy Survey).  Single power-law fits ($P$$\propto$$k^\gamma$) to 
the power spectra of the stronger-feedback (MaGICC) runs (over spatial 
scales corresponding to $\sim$0.5~kpc to $\sim$20~kpc) result in slopes 
consistent with those seen in the THINGS sample ($\gamma$$\sim$$-$2.5). The 
weaker-feedback (MUGS) runs exhibit shallower power law slopes 
($\gamma$$\sim$$-$1.2). The power spectra of the MaGICC simulations are more 
consistent though with a two-component fit, with a flatter distribution of 
power on larger scales (i.e., $\gamma$$\sim$$-$1.4 for scales in excess of 
$\sim$2~kpc) and a steeper slope on scales below $\sim$1~kpc 
($\gamma$$\sim$$-$5), qualitatively consistent with empirical claims, as 
well as our earlier work on dwarf discs. The radial HI surface density 
profiles of the MaGICC discs show a clear exponential behaviour, while those 
of the MUGS suite are essentially flat; both behaviours are encountered in 
nature, although the THINGS sample is more consistent with our stronger 
(MaGICC) feedback runs.
\end{abstract} 


\section{Introduction} 
\label{sec:Introduction}

The feedback of energy into the interstellar medium (ISM) is a fundamental 
factor in shaping the morphology, kinematics, and chemistry of galaxies, 
both in nature and in their simulated analogues \citep[e.g.][and references 
therein]{Thacker2000,Gov10,Schaye2010,Hambleton2011,Brook12, 
Scannapieco2012,Durier2012,Hopkins2013}.  Perhaps the single-most 
frustrating impediment to realising accurate realisations of simulated 
galaxies is the spatial `mismatch' between the sub-pc scale on which star 
formation and feedback operates, and the 10s to 100s of pc scale accessible 
to modellers within a cosmological framework. Attempts to better constrain 
`sub-grid' physics, on a macroscopic scale, have driven the field for more 
than a decade, and will likely continue to do so into the foreseeable 
future.

The efficiency and mechanism by which energy from massive stars (both 
explosive energy deposition from supernovae and pre-explosion radiation 
energy), cosmic rays, and magnetic fields couple to the ISM can be 
constrained indirectly via an array of empirical probes, including (but not 
limited to) stellar halo \citep{Brook2004} and disc \citep{Pilkington2012b} 
metallicity distribution functions, statistical measures of galaxy light 
compactness, asymmetry, and clumpiness \citep{Hambleton2011}, stellar disc 
age-velocity dispersion relations \citep{House2011}, rotation curves and 
density profiles of dwarf galaxies \citep{Maccio2012}, low- and 
high-redshift `global' scaling relations \citep{Brook12}, background QSO 
probes of the ionised circum-galactic medium \citep{Stinson2012}, and the 
spatial distribution of metals (e.g., abundance gradients and 
age-metallicity relations) throughout the stellar disc 
\citep{Pilkington2012a,Gibson2013}.
  
In \citet{Pilkington11}, we explored an alternate means by which to assess 
the efficacy of energy feedback schemes within a cosmological context: 
specifically, the predicted distribution of structural `power' encoded 
within the underlying cold gas of late-type \it dwarf \rm galaxies.  
Empirically, star forming dwarfs present steep spatial power-law spectra 
($P$$\propto$$k$$^\gamma$) for their cold gas, with $\gamma$$<$$-3$ on 
spatial scales $\simlt$1~kpc \citep{Stan99,Combes2012}, consistent with the 
slope expected when HI density fluctuations dominate the ISM structure, 
rather than turbulent velocity fluctuations (which dominate when isolating 
`thin' velocity slices).  Our simulated (dwarf) disc galaxies showed 
similarly steep ISM power-law spectra, albeit deviating somewhat from the 
simple, single, power-law seen by Stanimirovic et~al.

In comparison, the cold gas of late-type \it giant \rm galaxies appears to 
possess a more complex distribution of structural power. \citet{Dutta13} 
demonstrate that while such massive discs also present comparably steep (if 
not steeper) power spectra on smaller scales ($\gamma$$\sim$$-$3, for 
$\simlt$1kpc), there is a strong tendency for the power to `flatten' to 
significantly shallower slopes on larger scales ($\gamma$$\sim$$-$1.5, for 
$\simgt$2~kpc). Dutta et~al. propose a scenario in which the steeper power 
law component is driven by three-dimensional turbulence in the ISM on scales 
smaller than a given galaxy's scaleheight, while the flatter component is 
driven by two-dimensional turbulence in the plane of the galaxy's disc.

In what follows, we build upon our earlier work on dwarf galaxies 
\citep{Pilkington11}, utilising the Fourier domain approach outlined by 
\citet{Stan99}, but now applied to a set of four simulated massive 
($\sim$L$\star$) disc systems.  The simulations have each been realised with 
both conventional (i.e., moderate) and enhanced (i.e., strong/efficient) 
energy feedback.  The impact of the feedback prescriptions upon the 
distribution of power in the ISM of their respective neutral hydrogen (HI) 
discs will be used, in an attempt to constrain the uncertain implementation 
of sub-grid physics.  HI moment maps will be generated for each 
simulation \it and \rm (for consistency) massive disc from The HI Nearby 
Galaxy Survey (THINGS: \citealt{wal08}), to make a fairer comparison with the 
observational data. 

In \S\ref{sec:method}, the basic properties of the simulations 
are reviewed, including the means by which the HI moment 
maps, and associated Fourier domain power spectra, were analysed. The 
resulting radial surface density profiles, velocity dispersion profiles, 
and distributions of power in the corresponding cold interstellar media 
are described in \S\ref{sec:analysis}.  Our conclusions are presented in 
\S\ref{sec:conclusions}.

\section{Method}
\label{sec:method}

\subsection{Simulations}

Two $\sim$L$\star$ disc galaxies (\tt g1536\rm; \tt g15784\rm), 
drawn from the McMaster Unbiased Galaxy Survey (MUGS: \citealt{Stinson10}) 
and realised with the Smoothed Particle Hydrodynamics (SPH) code 
\textsc{Gasoline} \citep{Wad04}, form the primary inputs to our 
analysis.\footnote{The role of feedback in shaping the abundance gradients, 
metallicity distribution functions, and age-metallicity relations of these 
same four realisations has been presented recently by \citet{Gibson2013}.}
Two variants for each disc were generated, one employing
`conventional' feedback
(MUGS) and one using our 
`enhanced' feedback scheme (MaGICC: Making Galaxies In a Cosmological Context - 
\citealt{Brook12}; \citealt{Stinson2012}).\footnote{To link the 
simulation nomenclature with their 
earlier appearances in the literature, the MUGS variants of \tt g1536 \rm 
and \tt g15784 \rm are as first presented by \citet{Stinson10}, and analysed 
subsequently by \citet{Pilkington2012a} and \citet{Calura2012}, while the 
MaGICC variant of \tt g1536 \rm corresponds to the `Fiducial' run in 
\citet{Stinson2013} (itself, essentially the same as \tt SG5LR\rm, as first 
described by \citealt{Brook12}).}
These four massive disc simulations are referred to
henceforth as \tt g1536-MUGS, g1536-MaGICC, g15784-MUGS, \rm 
and \tt g15784-MaGICC, \rm and form the primary suite to which 
the subsequent analysis has been employed.
To provide a bridge to our 
earlier study of the ISM power spectra of dwarf galaxies
\citep{Pilkington11}, we have analysed an ancillary set of 
three simulated low-mass discs (\S3.4). \rm
An in-depth discussion of the MUGS and MaGICC star formation and feedback 
prescriptions are provided in the aforementioned works, although a brief 
summary of the key characteristics follows now.

The MUGS runs assume a thermal feedback scheme in which 4$\times$10$^{50}$ 
erg per supernova (SN) is made available to heat the surrounding ISM 
(`conventional'), while the MaGICC runs use 10$^{51}$ erg/SN (`enhanced').  
The MUGS simulations employ a \citet{Kroupa93} initial mass function (IMF), 
while MaGICC use the more `top-heavy' \citet{Chabrier2003} 
form.\footnote{The MUGS runs assumed that the global metallicity 
Z$\equiv$O+Fe, while those of MaGICC assume Z$\equiv$O+Fe+C+N+Ne+Mg+Si.}  
Radiation energy feedback from massive stars during their pre-SN phase 
(lasting $\sim$4~Myr) is included in the MaGICC runs, although it should be 
emphasised that the effective coupling efficiency is $<$1\% 
\citep{Brook12,Stinson2013}.  For both MUGS and MaGICC, cooling is disabled 
for gas particles situated within a blast region of size $\sim$100~pc, for a 
time period of $\sim$10~Myr. Star formation is restricted to regions which 
are both sufficiently cool and dense (MUGS: $>$1~cm$^{-3}$; MaGICC: 
$>$9~cm$^{−3}$). Metal diffusion \citep{SWS10} is included in all runs.

Supplementing the above four massive disc simulations, we have included
three lower mass dwarf discs: (a) SG2 and SG3 \citep{Brook12}
were realised with the same star formation and feedback schemes as the
MaGICC versions of g1536 and g15784, respectively; the only
difference lies in their initial conditions, where the former have been
`scaled-down' by an order of magnitude in mass; (b) DG1 \citep{Gov10}, 
the low mass
dwarf that formed the basis of our earlier work \citep{Pilkington11}.

\subsection{Analysis}
\label{subsec:analysis}

The analysis which follows is based upon a comparison of the HI gas 
properies of the MUGS+MaGICC simulations with their empirical `analogues', 
drawn from The HI Nearby Galaxy Survey (THINGS: \citealt{wal08}). We `view' 
the simulations face-on and restrict the comparison to massive discs from 
THINGS which are also close to face-on.  In practice, this has meant 
limiting the analysis to the same sub-sample as that used by 
\citet{Dutta13}. In contrast, our earlier work \citep{Pilkington11} focussed 
on low-mass dwarf galaxies, rather than massive discs; in that study, we 
found that the index of the simulated ISM power spectrum ($\gamma$, where 
$P$$\propto$$k^\gamma$) was consistent, to first order, with that observed 
in dwarfs (on spatial scales $\simlt$1~kpc) such as the Small Magellanic 
Cloud (i.e., $\gamma$$\sim$$-$3.2).  Besides determining the slope of the 
ISM power spectra for our new suite of massive disc galaxy simulations, we 
will present the radial HI surface density and velocity dispersion profiles, 
and contrast them with empirical data from the literature, in a further 
attempt to shed light on the role of feedback in shaping their 
characteristics.

In what follows, we make use of zeroth- (surface density) and second- 
(velocity dispersion) moment maps of each simulation's HI distribution 
(viewed, face-on), realised with the image processing package 
\textsc{Tipsy}.\footnote{\tt 
www-hpcc.astro.washington.edu/tools/tipsy/tipsy.html} The redshift $z$=0 
snapshots for each galaxy are first centred and aligned such that the angular 
momentum vector of the disc is aligned with the $z$-axis, and the neutral 
hydrogen fraction of each SPH particle inferred under the assumption of 
combined photo- and collisional-ionisation equilibrium. From the zeroth- 
(second-) moment maps, radial HI surface density (velocity dispersion) 
profiles were generated for each simulation and (near) face-on, late-type, 
disc from THINGS.  Individual results for each will be presented in 
\S\ref{sec:analysis}. It is worth noting that out of the THINGS galaxies 
presented in Figure~\ref{fig9}, NGCs~3031, 5236, 5457 and 
6946 are more extended than the VLA primary beam, resulting potentially in 
missing larger-scale information \citep{wal08}.

After \citet{Stan99} and \citet{Pilkington11}, the Fourier Transform of each 
of the aforementioned zeroth-moment HI maps (both simulations and empirical 
THINGS data) was taken, with circular annuli in Fourier space then employed 
to derive the average power in the structure of the ISM on different spatial 
scales.

\section{Results} 
\label{sec:analysis}

\subsection{Moment Maps}

\begin{figure*}
\centering
\psfig{file=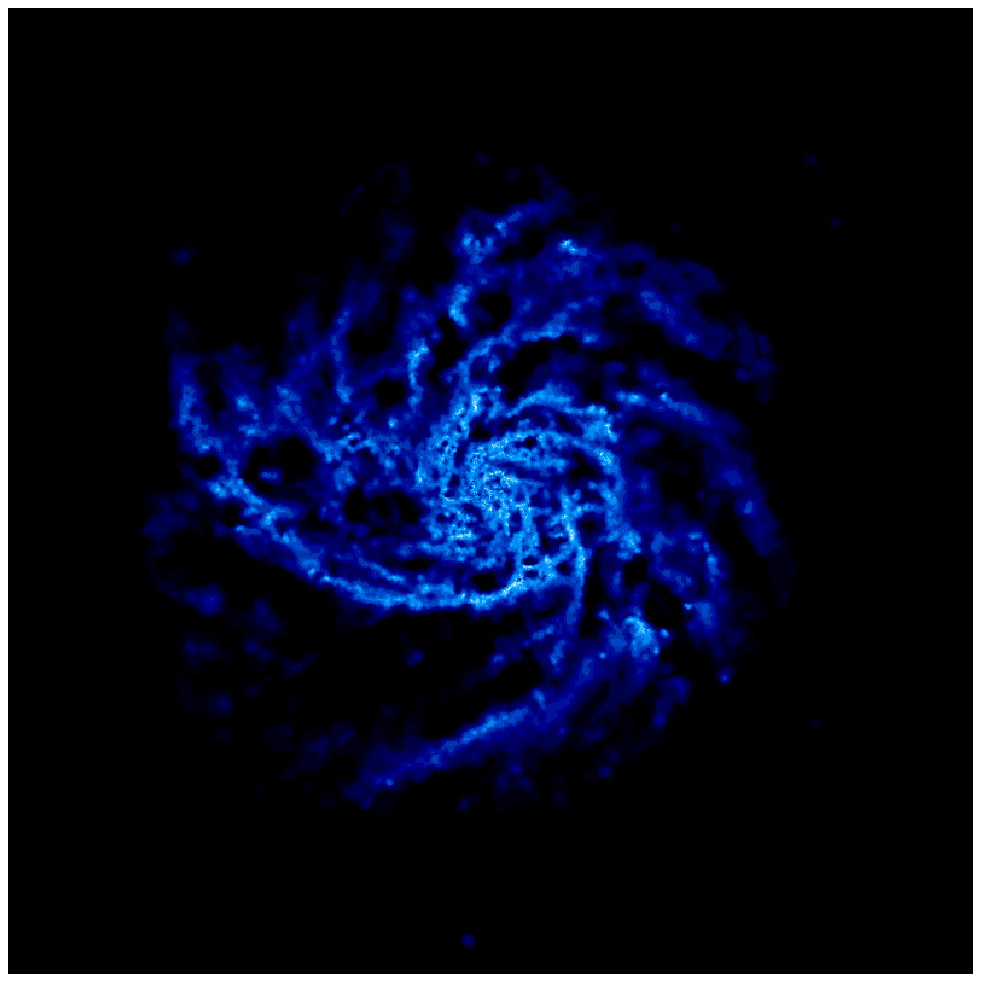,width=5.5cm}
\psfig{file=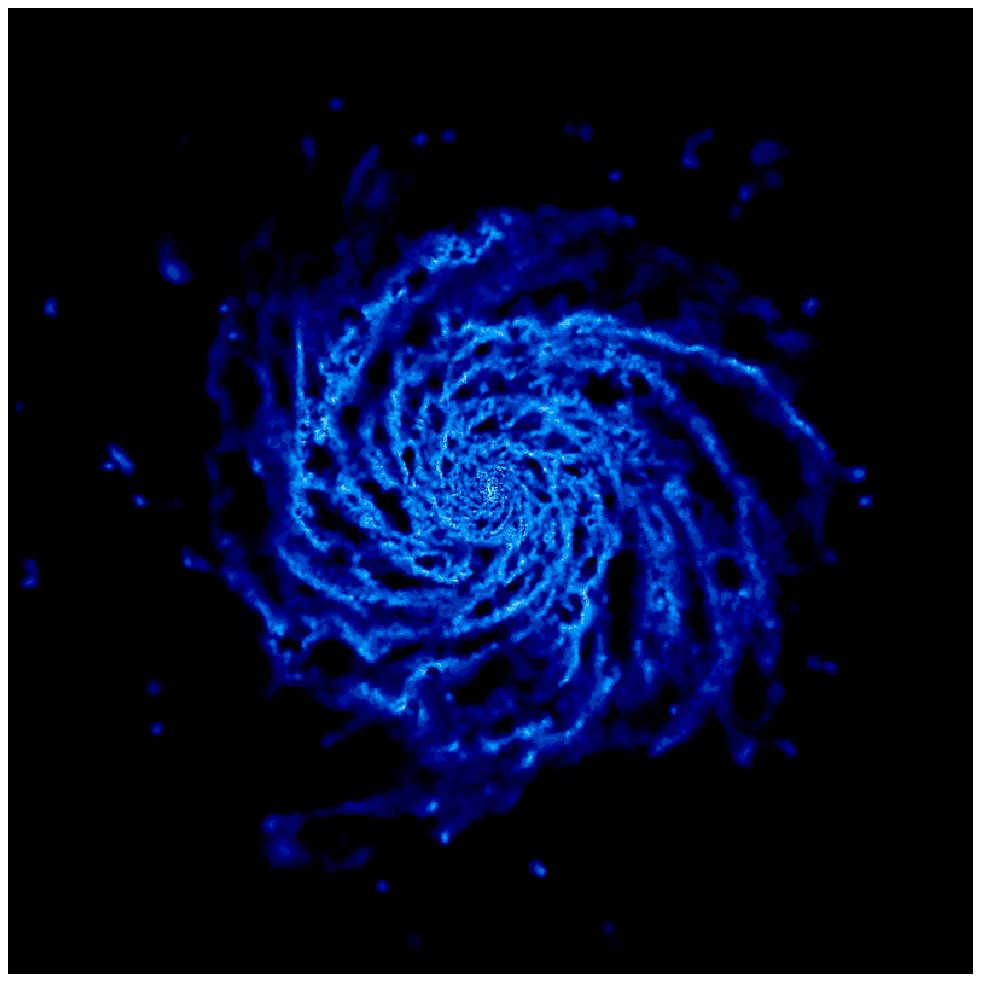,width=5.5cm}
\\
\psfig{file=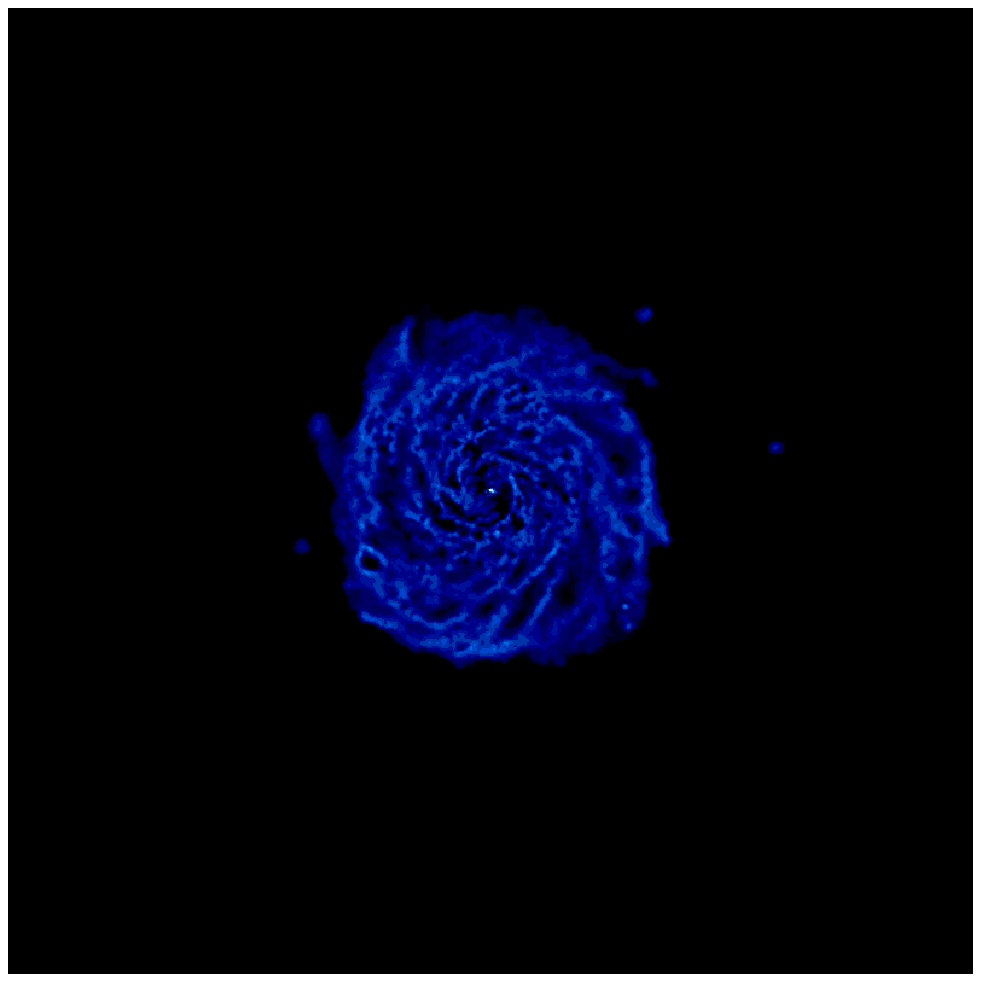,width=5.5cm}
\psfig{file=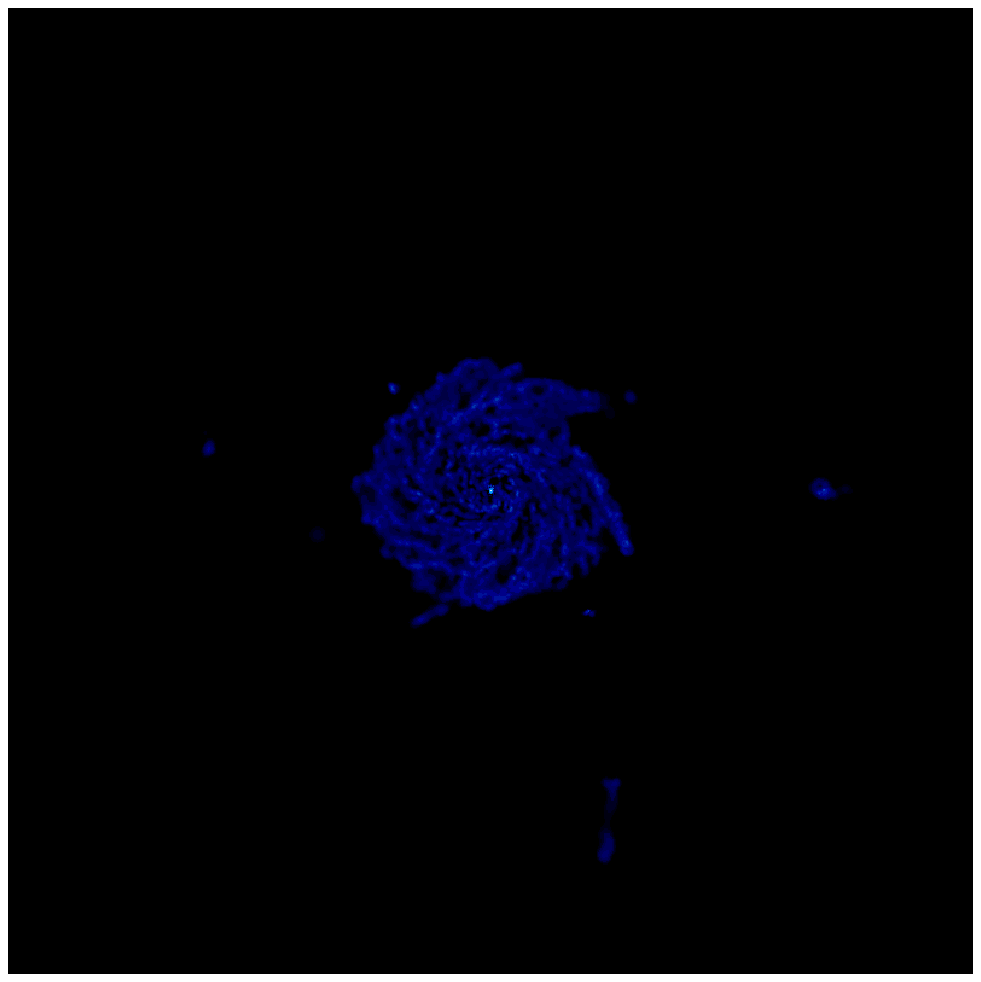,width=5.5cm}
\caption{Zeroth-moment HI maps for our four simulated $\sim$L$\star$ 
late-type discs: g1536-MaGICC (upper-left); g15784-MaGICC (upper-right); 
g1536-MUGS (bottom-left); g15784-MUGS (bottom-right). Each panel spans 
100$\times$100~kpc, with a column density range of 10$^{19}$~cm$^{-2}$ 
to 10$^{21}$~cm$^{-2}$ (comparable to the limits imposed by 21cm surveys 
such as THINGS).}
\label{fig1}
\end{figure*}

The zeroth-moment HI maps for our four simulated $\sim$L$\star$ late-type 
discs are shown in Fig~\ref{fig1}, with the two MaGICC (MUGS) variants shown 
in the upper (lower) panels.  Each panel spans 100$\times$100~kpc.  The 
`dynamic range' in HI column density in each panel is 
$\sim$10$^{19}$~cm$^{-2}$ to $\sim$10$^{21}$~cm$^{-2}$ - i.e., (roughly) the 
current observational lower and upper limits for HI (21cm) detection 
\citep{bigiel2008}.

Even a cursory inspection of Fig~\ref{fig1} suggests that the enhanced 
feedback employed within MaGICC results in significantly more extended HI 
discs, relative to the conventional feedback treatment within MUGS.  
Similary, at these column densities, the eye is drawn to the enhanced 
structure on larger scales seen in the MaGICC runs (relative to the more 
locally `confined' structure seen in MUGS).  Both points will be returned to 
below in a more quantitative sense.

\subsection{Radial Surface Density Profiles}

\begin{figure*} 
\centering
\psfig{file=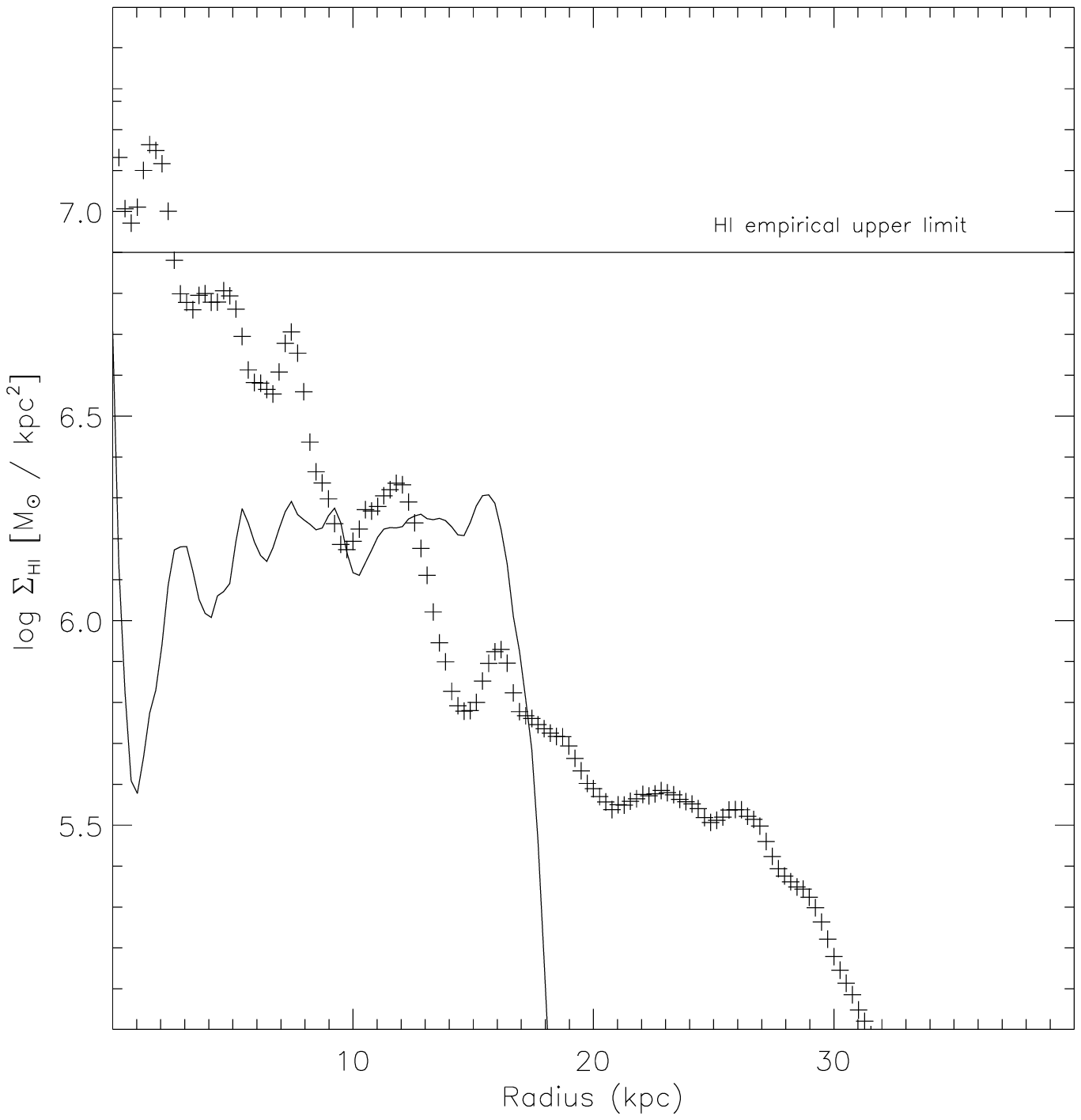, width=7.5cm}
\psfig{file=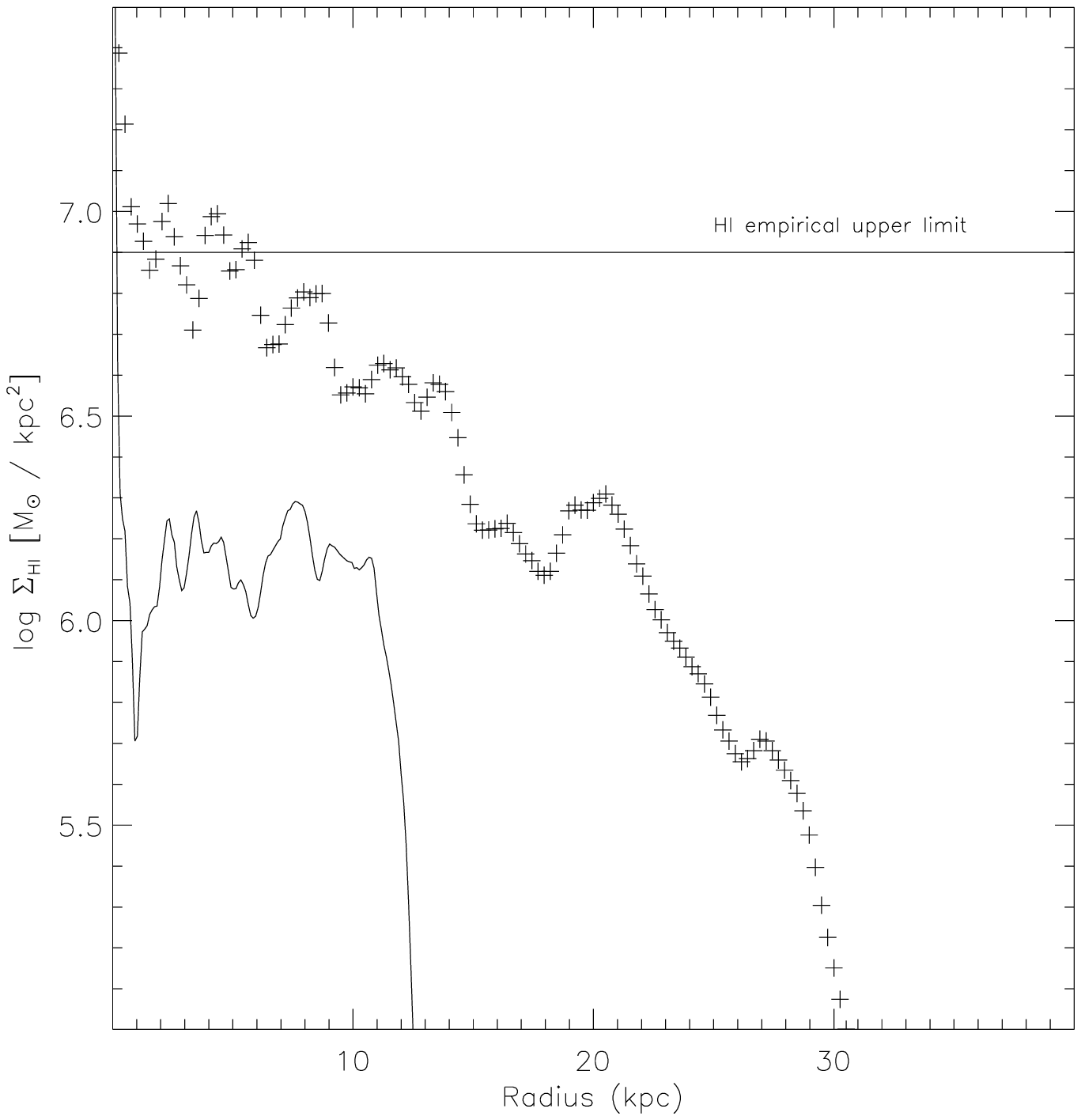,width=7.5cm}
\caption{Radial HI surface density profiles for the g1536 (left panel) 
and g15784 (right panel) simulations. The plus symbols represent the MaGICC 
runs and the solid lines correspond to the MUGS runs. The solid 
horizontal line in each panel correspond to the empirical HI upper limit
from \citet{bigiel2008}.} 
\label{fig2}
\end{figure*}

From the face-on moment zero maps of Fig~\ref{fig1}, radial HI surface 
density profiles were generated.  These are reflected in Fig~\ref{fig2} with 
the MUGS and MaGICC variants for g1536 (g15784) shown in the left (right) 
panel.  As for Fig~\ref{fig1}, the dynamic range has been limited to 
$\simgt$10$^{19}$~cm$^{-2}$ ($\simgt$10$^5$~M$_\odot$/kpc$^2$), to reflect 
the (typical) limiting 21cm detection limit in surveys such as THINGS; 
conversely, the horizontal line in each panel corresponds to the empirical 
HI upper limit (also from THINGS) of $\sim$10$^{6.9}$~M$_\odot$~kpc$^{-2}$.

The MaGICC discs (plus symbols in both panels) possess exponential 
surface density profiles (in HI) with $\sim$6$-$8~kpc scalelengths.
Conversely, the
MUGS realisations are clearly more `compact', with essentially `flat' 
radial HI surface density profiles (each with $\sim$10$^{20}$~cm$^{-2}$, 
independent of galactocentric radius), with an extremely `sharp' HI edge 
at $\sim$12$-$15~kpc.  At a limiting HI (21cm) column density of
$\sim$10$^{19}$~cm$^{-2}$, the MaGICC discs are $\sim$2$-$3$\times$
more extended than their MUGS analogues.
At first glance, in
terms of both radial dependence and amplitude, 
the HI surface density profiles of the MaGICC discs resemble very 
closely those shown in Fig~23 of O'Brien et~al (2010).
It is important to bear in mind though that
the O'Brien et~al. profiles were inferred (necessarily) from 
observations of edge-on discs.  Our analysis of the simulations
is restricted to face-on orientations, and so a fairer comparison
would be to the sample of \citet{Big&Blitz12}, who derived both
HI \it and \rm H$_2$ surface density profiles for a sample of 
face-on galaxies observed by THINGS.

\citet{Big&Blitz12} show that the HI in such disc galaxies is distributed more
uniformly, in terms of surface density, out to $\sim$10~kpc, with
(roughly) only a factor of $\sim$3 decline in going to a galactocentric
radius of $\sim$20~kpc.  This is consistent with the flatter gradient
seen for the MUGS simulations, albeit the issue of
their aforementioned overly 
truncated `edges' remains.
Because we cannot resolve the transition from HI to H$_2$ in our
simulations, some
fraction of what is labelled as `HI' in Fig~2 (at least within
the inner 5$-$10~kpc, for the MaGICC simulations, where the surface density 
is close to, or exceeds, the empirical upper limit for HI in nature) 
could certainly be misidentified H$_2$, and so our
inner gradients would be somewhat flatter than presented and 
therefore
more consistent with the profiles of Bigiel \& Blitz for radii
$\simlt$10~kpc. Our predicted HI surface density 
gradients in the $\sim$10$-$20~kpc range are (on average) somewhat steeper
than the typical galaxy from Bigiel \& Blitz (in the same 
radial range - see their Fig~1a), but certainly lie within $\sim$1$\sigma$
of the distribution. In that sense, the extended nature and (outer disc)
exponential profiles of the MaGICC simulations are more consistent with
those encountered in nature. \rm

\subsection{Radial Velocity Dispersion Profiles}

\begin{figure}
\centering
\psfig{file=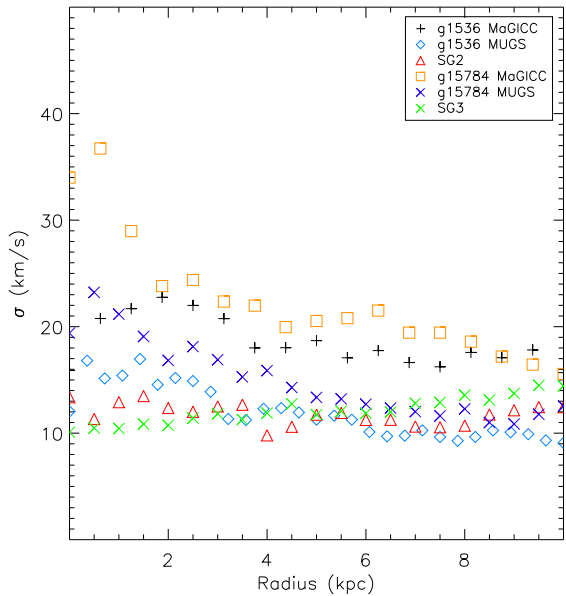,width=8.5cm}
\caption{Line-of-sight velocity dispersion profiles for the HI
within the face-on representations of g1536-MaGICC (black plus symbols);
g1536-MUGS (blue diamonds); g15784-MaGICC (red triangles); g15784-MUGS
(orange squares); SG2 (blue crosses), and SG3 (green crosses).} 
\label{figvel}
\end{figure}

The radial HI velocity dispersion profiles derived from the second-moment 
maps (Fig~\ref{figvel}) present fairly 'flat' trends with increasing 
galactocentric distance, save for perhaps g15784, with
$\sigma$ decreasing typically by $\sim$50\% in going from the 
inner disc to a galactocentric radius of $\sim$10~kpc; the profiles 
for the dwarfs (SG2 and SG3) are flat over this radial range, 
consistent with the dwarfs shown in Fig~3 of Pilkington et~al (2011).
Here, as the second-moment maps are for 
face-on viewing angles, the velocity dispersions quoted in Fig~\ref{figvel} 
are equivalent to $\sigma_{\rm W}$. We are only showing the velocity
dispersion profiles within the star-forming parts of the disks
(i.e., radii $\simlt$$r_{25}$ for the massive MaGICC and MUGS discs, 
and $\simlt$2$r_{25}$ for the lower mass dwarfs SG2 and SG3, where
$r_{25}$ is the isophotal radius corresponding to 25~mag~arcsec$^{-2}$).
As such, the dispersions being $\sim$20$-$100\% higher than the
`characteristic' value outside the star-forming disc ($\sim$10~km/s:
Tamburro et~al 2009) is not entirely unexpected.  

The thee main conclusions
to take from this part of the analysis are that: (i) the 
profiles and amplitudes
for the velocity dispersions of the cold gas \it within the
star-forming region \rm of the four massive MUGS and 
MaGICC discs overlap with those encountered in nature (Fig~1 of
Tamburro et~al.); (ii) the flat profiles of the two dwarfs (SG2 and SG3)
are more problematic, consistent with the conclusions of 
Pilkington et~al. (2011), and reflecting a limitation of our
inability to resolve molecular hydrogen processes on these scales;
(iii) the amplitudes
of the MaGICC variants, relative to their MUGS counterparts, are $\sim$50\% 
higher (although both are within the range encountered in nature); 
such a 
result is not entirely unexpected, given the significantly enhanced feedback 
associated with the MaGICC runs.

\subsection{Power Spectra}

As noted in \S\ref{subsec:analysis}, power spectra were derived from each of 
the simulated and empirical (THINGS) HI moment-zero maps, by averaging in 
circular annuli in frequency space after Fourier transforming the images.  
The technique is identical to that employed by \citet{Stan99} and 
\citet{Pilkington11}. While alternate approaches certainly exist (cf. 
\citealt{Dutta13}), we are more concerned here with adopting a homogeneous 
approach for both the simulations and the data, rather than necessarily 
inter-comparing the various techniques available.

\begin{figure} 
\centerline{ 
\psfig{file=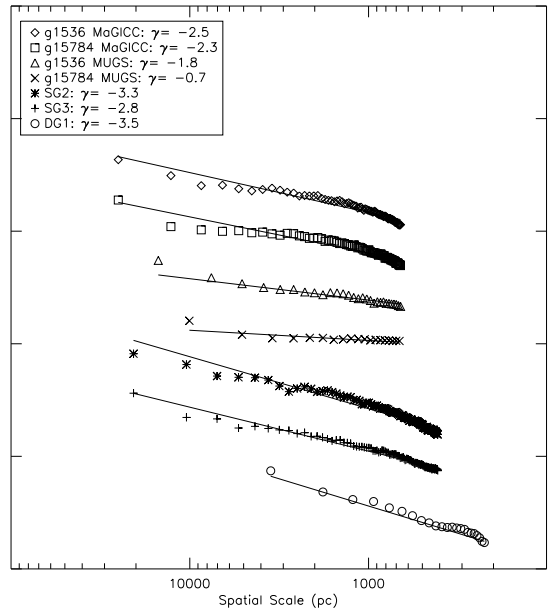,width=8.5cm}
} 
\caption{Power spectra for the four $\sim$L$\star$ MaGICC and MUGS 
simulations (upper four spectra), two `dwarf' variants of g1536 and 
g15784 (SG2 and SG3, respectively), and the low-mass dwarf DG1, from 
\citet{Pilkington11}. The inset within the panel links the symbol with 
the relevant simulation.  The ordinate represents arbitrary units of 
spatial power, as the relative distribution (rather than absolute) is 
the focus of this work; each spectrum has been offset with respect to 
the next, for ease of viewing.}
\label{fig5}
\end{figure}  

Fig~\ref{fig5} shows the power spectra for both the MaGICC and MUGS variants 
of g1536 and g15784 simulations, as well as their respective dwarf galaxy 
analogs, SG2 and SG3.  For each of the four massive
discs' spectra, single power-law fits are shown (solid curves) for the spatial 
scales over which the fits were derived ($\sim$0.6$-$2~kpc). It should be 
emphasised that the lower limit on the spatial scale over which these fits 
were made corresponds to twice the softening length employed in the 
simulations; while an argument could be made to extending to somewhat 
smaller scales, we felt it prudent to be conservative in our analyses. What 
can hopefully be appreciated from a cursory analysis of Fig~\ref{fig5} is 
the relatively enhanced power on sub-kpc scales seen in MUGS (conventional 
feedback) realisations, compared with the their MaGICC (enhanced feedback) 
analogues.  This is reflected in the single power-law slopes itemised in the 
inset to the panel (which are weighted heavily by the more `numerous' higher 
frequency `bins' on sub-kpc scales), which are meant to be illustrative 
here, rather than represent the formal `best fit' to the data.  Broadly 
speaking, the power spectra are roughly an order of magnitude steeper when 
using the MaGICC feedback scheme, as opposed to that of MUGS - i.e., \it it 
appears that the stronger feedback shifts the ISM power from predominantly 
`small' ($\simlt$1~kpc) to large ($\simgt$2~kpc) spatial scales. \rm

We next extended our analysis to lower mass, late-type, systems, including 
the two dwarf variants to g1536-MaGICC and g15784-MaGICC (referred to as SG2 
and SG3, as per \citealt{Brook12}).  We also performed an independent 
re-analysis of the dwarf (DG1) that formed the basis of our earlier work 
\citep{Pilkington11}.  The inclusion of these three `dwarfs' allows us to 
push the analysis to somewhat smaller spatial scales, while still working 
within a framework of `enhanced' feedback.  The power spectra for all seven 
systems are shown in Fig~\ref{fig5}.  An important conclusion to be drawn 
from this figure (and associated quoted single power-law fits within the 
inset to the panel) is that \it on $\sim$sub-kpc scales, the power spectra 
slopes of the three dwarfs (SG2, SG3, DG1) are steeper 
($-$3.5$\simlt$$\gamma$$\simlt$$-$3) then their more massive
analogues. \rm

\begin{figure} 
\centerline{ 
\psfig{file=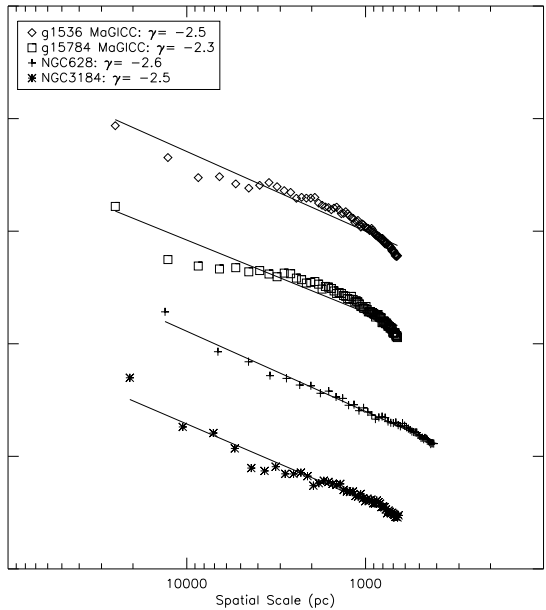,width=8.5cm}
} 
\caption{Power spectra for the two MaGICC simulations, and two selected 
from the empirical THINGS dataset (NGC~628 and 3184). All other details 
are as per the caption to Fig~\ref{fig5}.}
\label{fig4}
\end{figure} 

We then compared the predicted power spectra from the two $\sim$L$\star$ 
discs realised with the enhanced MaGICC feedback scheme, with those derived 
from galaxies from the THINGS database; the full database is shown in 
Fig~\ref{fig9}, but for succinctness, we only show the power spectra for 
NGC~628 and 3184 (which were chosen, in part, because they were the closest 
to face-on, matching, by construct, the MaGICC simulations), alongside the 
MaGICC discs, in Fig~\ref{fig4}.  In terms of formal single power-law fits 
to these spectra, the MaGICC and (selected) THINGS galaxies are very similar 
(as shown by the quoted slopes within the inset to the panel).  Having said 
that, as already alluded to in relation to Fig~\ref{fig5}, the MaGICC 
spectra do not appear entirely consistent with a single power-law, instead 
presenting evidence for something of a `break' in the structural power, on 
the scales of $\sim$1$-$2~kpc (being flatter on larger scales, and steeper 
on smaller scales, a point to which we return below).

\begin{figure*} 
\centering
\psfig{file=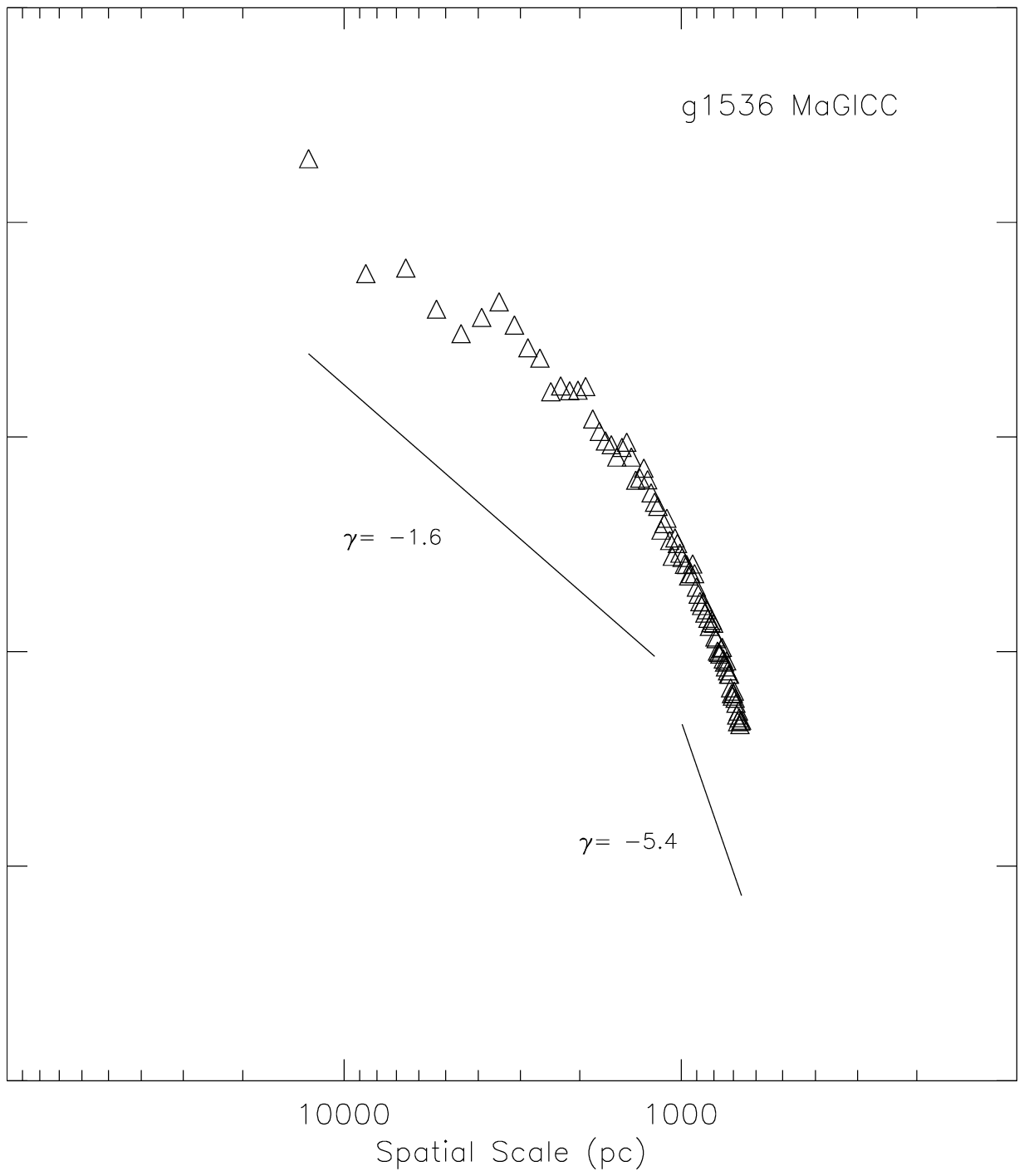,width=5.0cm}
\psfig{file=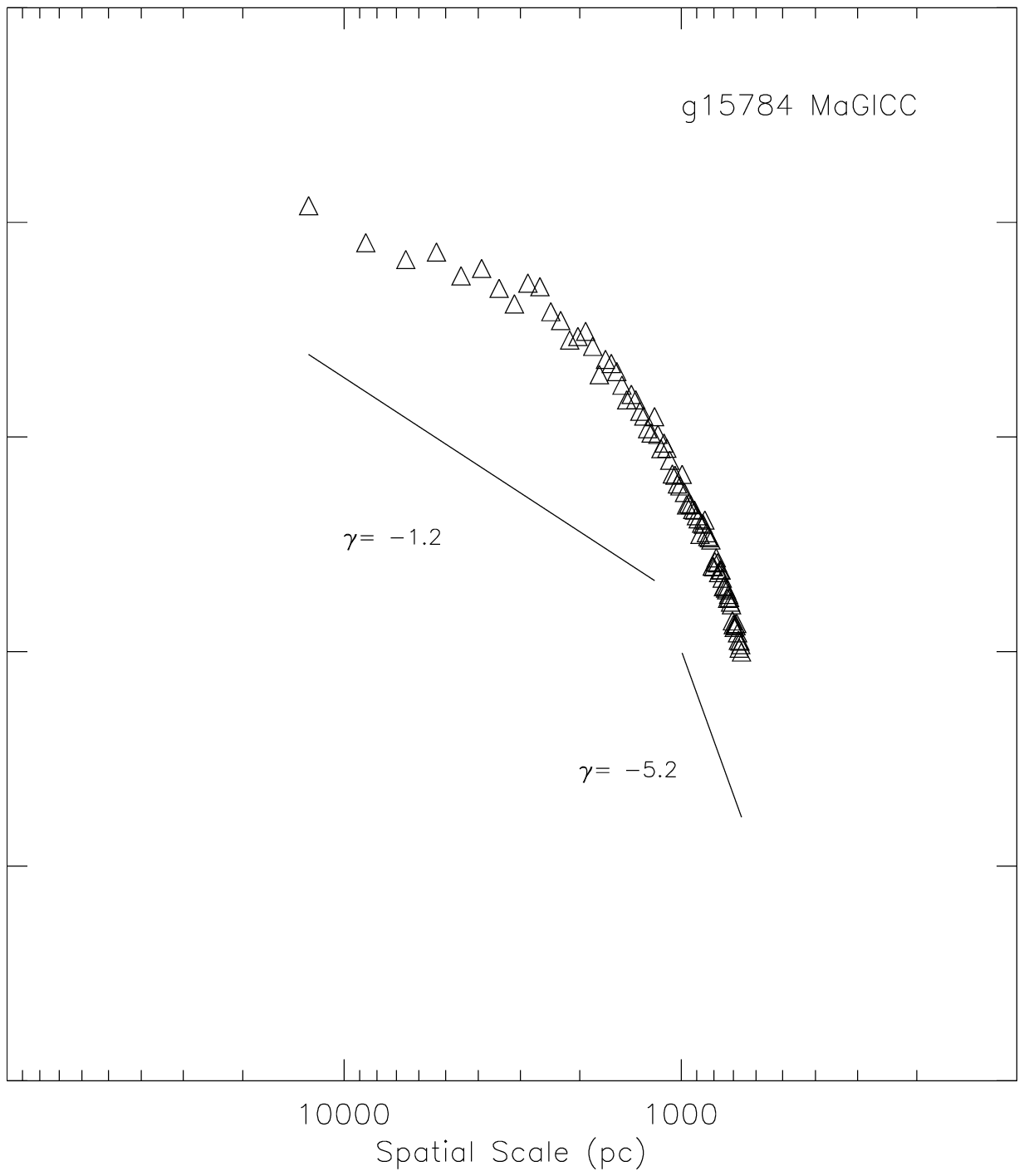,width=5.0cm}
\psfig{file=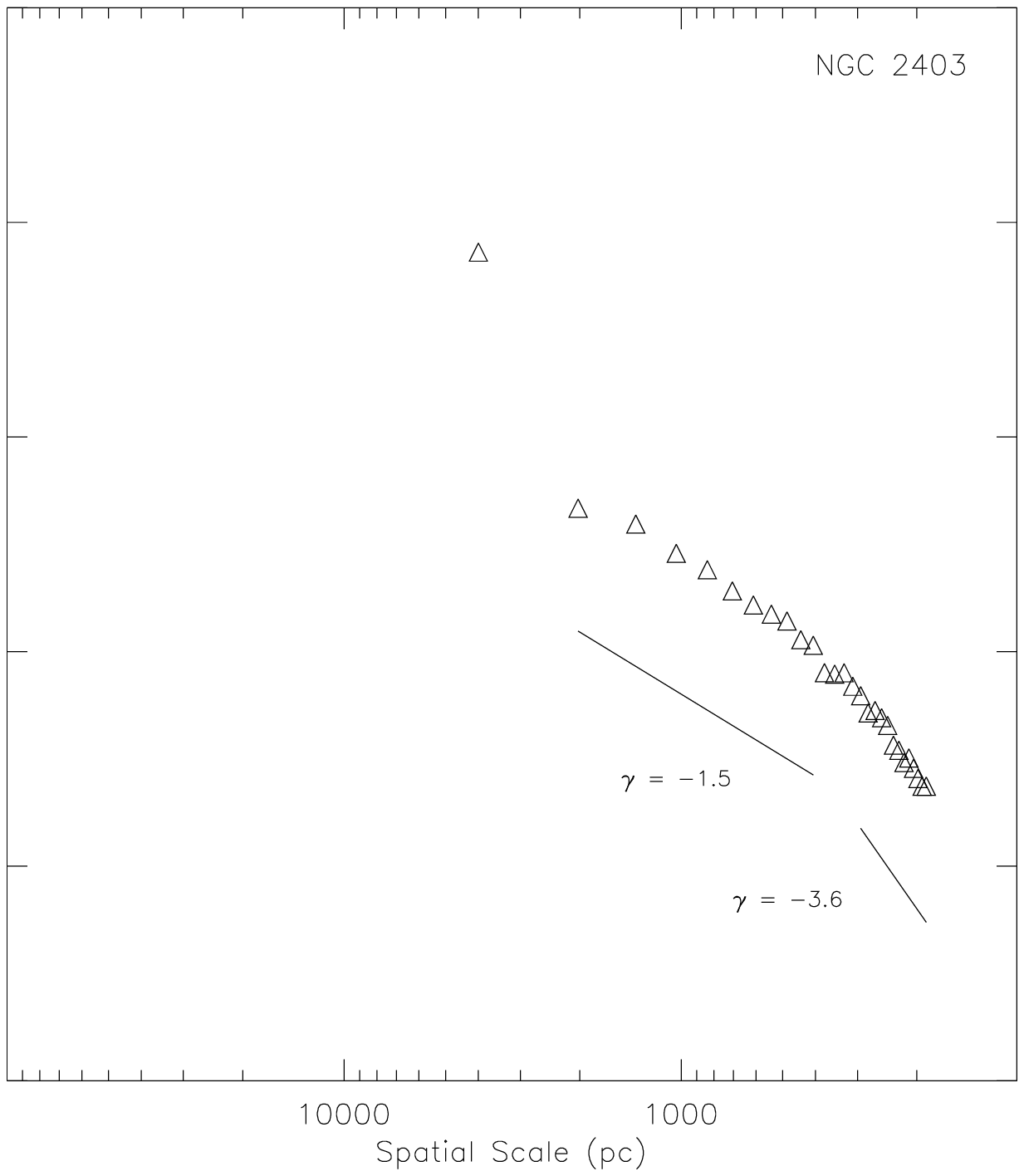,width=5.0cm}
\caption{Power spectra of g1536-MaGICC, g15784-MaGICC, and NGC~2403, 
respectively (from left to right). Each spectrum appears (in a 
qualitative sense) inconsistent with a single power-law fit; 
two-component fits, with a shallower (steeper) slope on larger (smaller) 
scales, are suggested, although the `knee' in the spectra occurs on 
different scales for the simulations ($\sim$2~kpc), as opposed to that 
of NGC~2403 ($\sim$0.5~kpc).}
\label{fig6}
\end{figure*}

Inspection of Figs~\ref{fig5} and \ref{fig4} suggests that single power law 
fits are not necessarily the best option.  In Fig~\ref{fig6}, we show the 
result of performing two-component fits to both the MaGICC data and a 
selected galaxy from THINGS (NGC~2403, chosen as it is the THINGS galaxy 
whose power spectrum looks like it would suit a 2-component fit best).  In a 
qualitative sense, the behaviour is not dissimilar - i.e., both the MaGICC 
simulations and NGC~2403 show flatter power spectra on larger scales, 
compared with smaller scales, although the transition from `flat' to `steep' 
occurs at $\sim$2~kpc in the simulations, as opposed to $\sim$0.5~kpc in 
NGC~2403. This seems to be consistent with the idea posed by \citet{Dutta13} 
that there is a steep power-law component on smaller scales driven by 
3-dimensional turbulent motions, which flattens at larger spatial scales. At 
these larger scales, 2-dimensional turbulent motions begin to dominate 
within the plane of the galactic disc. 
The steepening of the power spectra on small spatial scales observed in 
the power spectra of the MaGICC large discs is also seen in work undertaken 
by \citet{Elmegreen01} in their work on the LMC.

Power spectra have been generated for the 17 
THINGS galaxies employed in the analysis of \citet{Dutta13}; these are 
provided in the accompanying Appendix as Fig~7.  
The majority have slopes on 
the order of $\gamma$$\sim$$-$2.3 to $-$2.8, with two exceptions: NGC~3031 
($\gamma$$\sim$$-$0.9) and NGC~3521 ($\gamma$$\sim$$-$3.3). Much as for the 
simulations, the point associated with the largest spatial scales in each 
panel should be viewed with some skepticism, as edge effects do come into 
play (i.e., the `edge' of the HI disc is `seen' as a high power `scale' 
against an almost noise-free background).

\section{Conclusions} 
\label{sec:conclusions}

We have presented an analysis of the cold gas and HI content of simulated 
discs with both 'standard' (MUGS) and `enhanced' (MaGICC) energy feedback 
schemes, as well as re-scaled dwarf variants of the massive (MaGICC) 
simulations.

Radial density profiles were generated for the MUGS and MaGICC 
$\sim$L$\star$ variants of \tt g1536 \rm and \tt g15784 \rm (Fig~2). These 
were generated using their respective zeroth HI moment maps; the weaker 
feedback associated with MUGS resulted in very flat radial HI distributions, 
with sharp cut-offs at galactocentric radii of $\sim$12$-$15~kpc, while the 
stronger feedback associated with MaGICC resulted in HI discs with 
exponential surface density profiles (with scalelengths of $\sim$6$-$8~kpc) 
which were $\sim$2$-$3$\times$ more extended (at an HI column density limit 
of $\sim$10$^{19}$~cm$^{-2}$).  The exponential profiles exhibited by the 
enhanced feedback runs are consistent with the typical profile observed in 
nature \citep{bigiel2008,Obr10}. The majority of the THINGS radial density 
profiles show evidence of exponential components, indicating that the MaGICC 
simulations distribute the column density in a way that better matches 
observational evidence.

The power spectra generated for the massive ($\sim$L$\star$) discs with 
enhanced (MaGICC) feedback are steeper than their weaker (MUGS) feedback 
counterparts. In other words, the stronger feedback shifts the power in ISM 
from smaller scales to larger scales.  Forcing a single component power-law 
to the MaGICC spectra yields slopes consistent with similarly forced single 
component fits to the empirical THINGS spectra. also well-described by a 
single component power law; having said that, the MaGICC spectra are more 
consistent with a two-component structure, with a steeper slope on sub-kpc 
spatial scales, flattening to shallower slopes on larger scales.  The 
massive discs realised with the MUGS feedback scheme are both shallower than 
MaGICC, but also well-fit with a single power-law across all spatial scales. 
The dwarf galaxies realised in our work with enhanced feedback possess 
steeper slopes than their more massive counterparts, with values that are in 
agreement with \citet{Stan99} and \citet{Pilkington11}.

It is arguable that several of the THINGS power spectra warrant 
multiple-component fits (namely NGC 2403, 3031, 3184, 3198 and 7793) and the 
multi-component fits performed on NGC 2403 and the two large disc MaGICC 
galaxy power spectra indicate that the large-scale slopes agree well, 
whereas the small scale slopes differ largely. This indicates that the 
MaGICC feedback scheme distributes HI structures on a scale that is 
comparable to those of observational results, but there is a lack of 
small-scale structure. It is apparent that there is no 1:1 match to the 
THINGS data from either the MUGS or MaGICC feedback schemes, but MaGICC 
appears to fare better than the MUGS feedback scheme from a single-component 
fit in an average sense. The lack of a 1:1 relation may be largely due to 
the challenges in converting from 'cold gas' to 'HI' as well as a lack of 
exactly face-on systems observed in nature and in the THINGS survey. 

\section*{Acknowledgments} 

BKG acknowledges the support of the UK’s Science \& Technology Facilities 
Council (ST/J001341/1). KP acknowledges the support of STFC through its PhD 
Studentship programme (ST/F007701/1). The generous allocation of resources 
from STFC’s DiRAC Facility (COSMOS: Galactic Archaeology),
the DEISA consortium, co-funded through EU FP6 
project RI-031513 and the FP7 project RI-222919 (through the 
DEISA Extreme Computing Initiative), the PRACE-2IP Project (FP7 RI-283493),
and the University of Central 
Lancashire’s High Performance Computing Facility.

\bibliographystyle{mn2e} 
\bibliography{PowSpectra} 

\section*{Appendix}

We present here the ISM power spectra for 
the 17 THINGS galaxies used in this work. The inset to each panel includes 
the galaxy name, the weighting scheme employed (RO=robust), and the best-fit 
(single component) power-law slope.

\begin{figure*}
\centering
\psfig{file=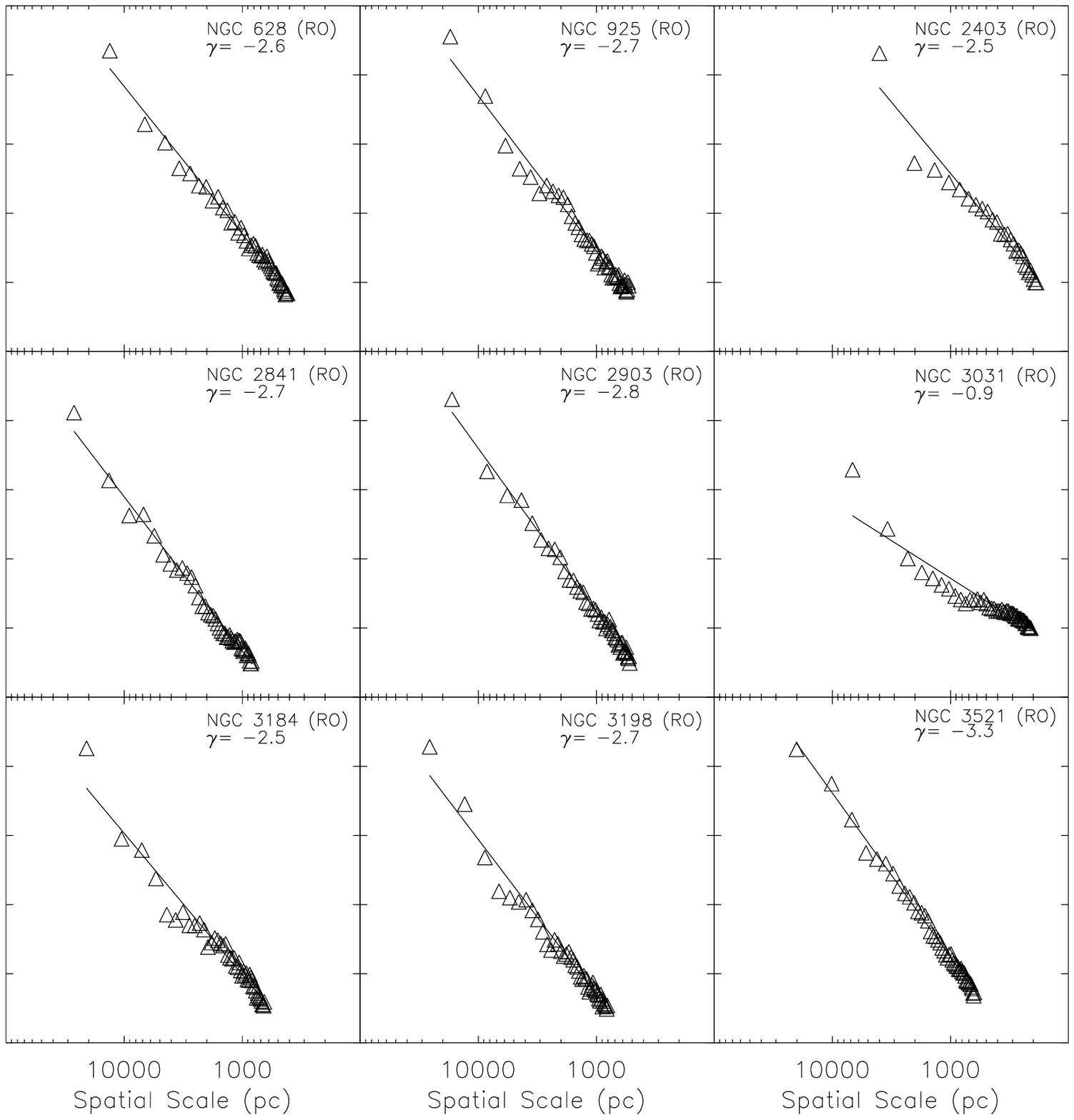,width=9.0cm}
\\
\psfig{file=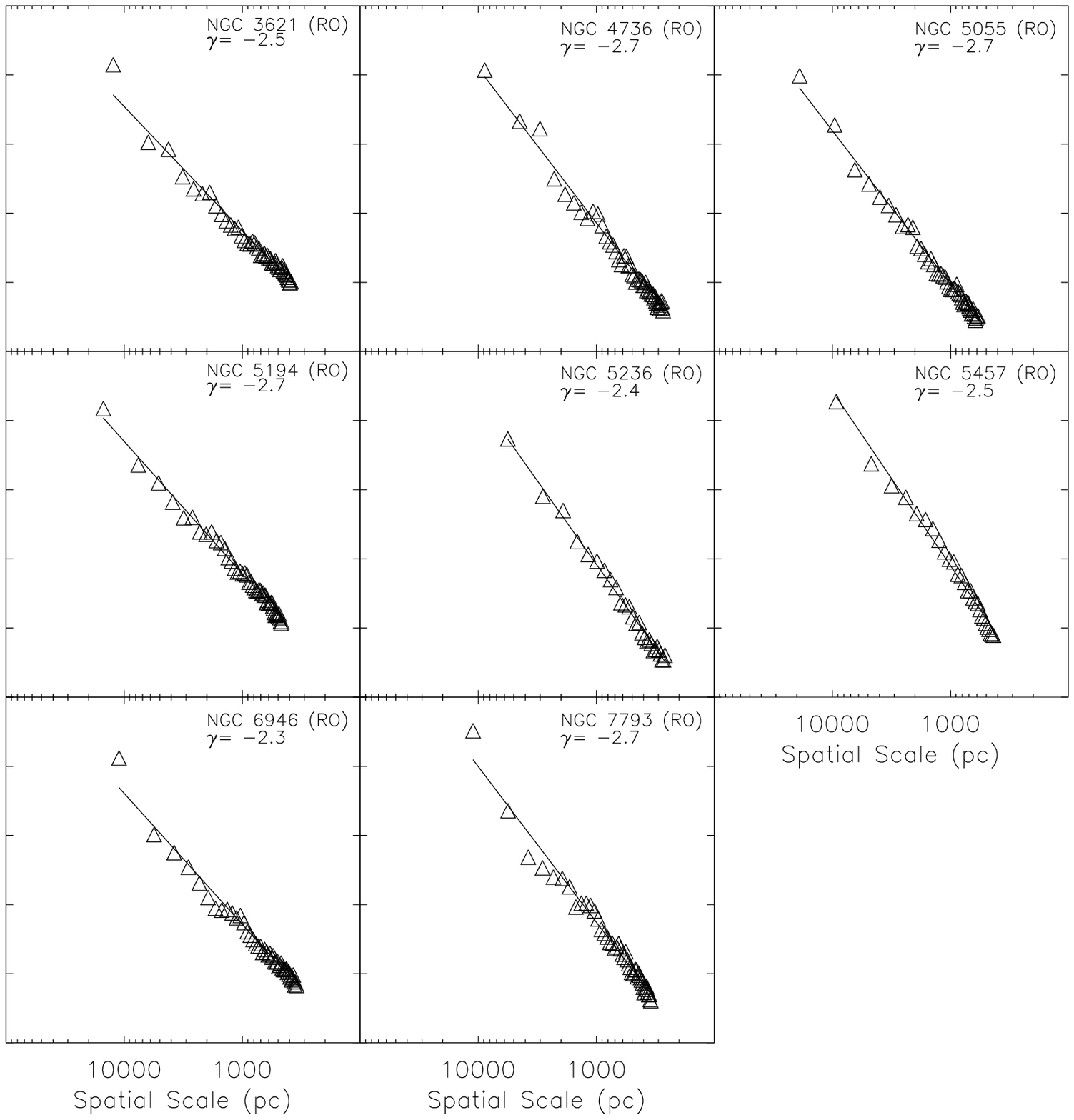,width=9.0cm}
\caption{Power spectra for all the THINGS galaxies analysed in this work; names of the galaxies are listed on their corresponding plots along with the power law slope value. The power law slope is plotted over the points as a solid line.}
\label{fig9}
\end{figure*}

\label{lastpage}

\end{document}